\documentclass{emulateapj}

\usepackage{epstopdf}

\newcommand{\gray}{$\gamma$-ray\ } \newcommand{\grays}{$\gamma$-rays\ }

\begin{document}
\DeclareGraphicsExtensions{.pdf,.gif,.jpg}

\shorttitle{unresolved EGB from Core Dominated RGs}
\shortauthors{TBD}

\title{The Extragalactic Gamma-Ray Background from Core Dominated Radio Galaxies}

\vspace{2cm}

\author{
Floyd W. Stecker\altaffilmark{1,2},
Chris R. Shrader\altaffilmark{1,3},
Matthew. A. Malkan\altaffilmark{2}}

\altaffiltext{1}{Astrophysics Science Division, NASA/Goddard Space Flight Center, Greenbelt, MD 20771}
\altaffiltext{2}{Department  of Physics  and  Astronomy, University of California, Los Angeles,
Los   Angeles,  CA 90095-1547}
\altaffiltext{3}{Physics Department, Catholic University of America, Washington, DC 20064}

\begin{abstract}

Recent radio surveys have discovered a large number of low luminosity core dominated radio galaxies that are much more abundant than those at higher luminosities. These objects will be too faint in \grays to be detected individually by {\it Fermi}. Nevertheless, they may contribute significantly to the unresolved extragalactic \gray background. We consider here the possible contribution of these core dominated radio galaxies to the diffuse extragalactic \gray background. Using published data available for all 45 of the radiogalaxies listed as detected counterparts in the {\it Fermi} FL8Y source list update to the 3FGL catalog, we have searched for radio maps which can resolve the core flux from the total source flux. Using high resolution radio maps we were able to obtain core fluxes for virtually every source. We then derived a relation between core radio flux and \gray flux that we extrapolated to sources with low radio luminosities that are known to be highly core dominated. We then employed a very recent determination of the luminosity function for core dominated radio galaxies in order to obtain the contribution of all possible \gray emitting radio galaxies to the unresolved extragalactic \gray background. We find this contribution to be a possibly non-negligible, $4\% - 18\%$ of the unresolved \gray background observed using the {\it Fermi}-LAT telescope.
 
\end{abstract}


\section{Introduction} 

The Large Area Telescope on the {\it Fermi} Gamma-ray Space Telescope (hereafter called {\it Fermi}) has measured the diffuse {\it unresolved} extragalactic \gray background (hereafter, EGB) at energies between 0.1 GeV and 820 GeV (Ackermann et al 2015). The photon spectrum of the EGB is found to follow an approximate power-law of spectral index $\sim$ 2.3 below $\sim$ 200 GeV, with a total flux above 0.1 GeV of $\sim 7 \times 10^{-6}$ cm$^{-2}$ s$^{-1}$ sr$^{-1}$. It appears that $\sim$ 85\% of the EGB at energies above 50 GeV is from unresolved blazars (Ackermann et al. 2016). Below 50 GeV, the EGB may be made up of various unresolved extragalactic \gray sources such as blazars (e.g., Stecker 1993; Padovanni et al. 1993; Stecker \& Salamon 1996; Stecker \& Venters 2011) and star forming galaxies (Pavlidou \& Fields 2001,2002; Stecker 2007; Fields 2010; Makiya et al. 2011; Stecker \& Venters 2011; Tamborra et al. 2014; Linden 2017). Star forming galaxies emit \grays at energies above 0.1 GeV, such \grays being produced by interactions of cosmic rays accelerated in supernova explosions of young massive stars with interstellar gas (Stecker 1971, 1976). If one lists the number of \gray point sources by AGN type one finds that the gamma-ray sky is dominated by radio-loud blazars that have their jets aligned within $\sim 20^{\circ}$ of the direction of the Earth. No radio quiet AGN, other than nearby ones with active star formation, have been detected by  {\it Fermi}. 

In addition to blazars and star forming galaxies, a number of \gray emitting radio galaxies have been detected by {\it Fermi}, indicating that radio galaxies can also contribute to the EGB (e.g., Inoue 2012; di Mauro et al. 2014; Hooper et al. 2016). These galaxies are believed to be AGN whose jets are "mis-aligned" with the direction of the Earth (Inoue 2011).\footnote{There is no standard definition of "mis-alligned". As opposed to blazars (see above), we can take this criterion to be $\sim > 20^{\circ}$ of the direction of the Earth.}
In this study, we specifically consider the more recently discovered relatively abundant classes of low luminosity core dominated radio galaxies. We thus make here the hypothesis that low luminosity core dominated radio galaxies may contribute significantly to the unresolved extragalactic \gray background. In order to examine this hypothesis we derive a new relation between core radio and \gray luminosity based on an updated list of {\it Fermi}-detected radio galaxies. This relation implies that lower luminosity \gray emitting radio galaxies are expected to have proportionally lower \gray luminosities so that they cannot be detected individually by {\it Fermi}. However, their high relative abundance implies that they can contribute to the unresolved (diffuse) \gray background. In order to examine this hypothesis we need to derive a new relation between core radio and \gray luminosity.

In Section \ref{newsources} we discuss how we determined estimated core radio fluxes from radio maps for the \gray emitting radio sources given in the {\it Fermi} FL8Y source list. The FL8Y source list, which is based on the first eight years of {\it Fermi} data in the 100 MeV - 1 TeV energy range, includes 5523 sources observed by {\it Fermi} as having a significance $> 4\sigma$. 

In section \ref{rl-gl} we use the expanded {\it Fermi} FL8Y source list, together with the results from Section \ref{newsources}, to derive a new log-linear relation between \gray luminosity and core radio luminosity for {\it Fermi}-detected radio galaxies. 

In Section \ref{lowlum} we make the case for the importance of the potential contribution of classes of low luminosity radio sources to the diffuse EGB; such sources being too weak to be individually detected by {\it Fermi}. 

Section \ref{LF} gives our choice for  luminosity functions for core dominated radio galaxies at different redshifts. This includes the
classes of low luminosity radio sources considered in Section \ref{lowlum}. 

Section \ref{EGB} outlines our calculation of the EGB from \gray emitting radio galaxies. Section \ref{concl} concludes with a discussion of our results.

\section{Core Luminosities for our Data Set of Gamma-ray Emitting Radio Galaxies}
\label{newsources}

As a first step in our program to determine the EGB from core dominated radio galaxies, we
examined the published radio and \gray data available for all 45 of the radio galaxies listed as detected counterparts in the {\it Fermi} FL8Y source list update to the {\it Fermi} 3FGL source catalog \footnote{\small{The FL8Y list is based on 8 years of sky survey data. It is a precursor to the forthcoming 4FGL catalog.}} 
\footnote{\tt https://fermi.gsfc.nasa.gov/ssc/data/access/lat/fl8y/}. We employed this updated list of 45 sources, in conjunction with their corresponding radio data discussed in Section \ref{newsources}, to derive a new radio (core) to gamma-ray luminosity relation. 

In order to obtain the core fluxes for these 45 sources, we searched for radio maps of these sources which can resolve the core flux from the total source flux. For virtually every source we were able to find these measurements. We sought radio observations the highest angular resolution available. In the majority of cases this was provided by interferometry at the Very Large Array (VLA).\footnote{\small{For those sources which are too far south for the VLA to observe, we found maps from the Australia Telescope Compact Array (ATCA). We were also able to obtain results from Multi-Element Radio Linked Interferometer Network (MERLIN).}} Nearly all of the sources have a detectable compact core with a measured flux. The core observed fluxes range from making up $>$ 90\% of the total source flux, down to 10 \% or less. 

In a few cases, no distinct 'core' component is visible. For such radio galaxies we
used radio maps to set strict upper limits to the possible core flux, which is typically
5\% or less of the total flux.\footnote{\small{Only 3 of the Fermi-detected radiogalaxies show no detectable cores, only diffuse extended (usually double) radio emission: and MRC0522-239, MRC1303-215, and MRC 2327-215. For these we adopted conservative upper limits to the possible flux of a faint core.}} Most of the maps were made at an observing frequency of 5 GHz. Where these were not available, we used 1.4 GHz maps. References to each of the publications used to determine the core fluxes are given in Table 1 and more fully in the reference list.

\section{Radio Luminosity - Gamma Ray Luminosity Correlation}
\label{rl-gl}

We now derive a relationship between the expected \gray luminosity of a radio galaxy and its radio luminosity. Fortunately, such a relationship has been found to exist that spans the entire radio luminosity range, regardless of type (such as FRI and FRII), provided one considers only the {\it core luminosity} of the radio sources (here designated by $L_R$) rather than their total luminosity (Inoue 2011; di Mauro et al. 2014; Hooper et al. 2016). This relation can be understood conceptually. The radio core emission is directly associated with the power produced by non-thermal processes in the central engine of the AGN in the vicinity of the black hole and the base of the jet. The physics of such jets is analogous to that of the more familiar \gray emitting blazars such as BL Lac objects. In fact, it appears that {\it Fermi}-detected \gray emitting blazars are more core dominated than those not detected by {\it Fermi} (Pei et al. 2018). 

We have updated the relation between \gray luminosity and core radio luminosity using our expanded data set as discussed in the previous section. We therefore will adopt the reasonable assumption of extrapolating our new relation between radio core and \gray luminosities to lower core radio luminosities.

For \gray emitting radio galaxies, di Mauro et al. (2014) and Hooper et al. (2016) found a relation between the radio (core) and \gray  luminosities. Those authors found an approximate log-linear relation to hold,
i.e. 
\begin{equation}
\log(L_\gamma) = A \log (L_R) + B.
\label{lum-lum}
\end{equation}
Subsequent to those previous determinations, the total {\it Fermi} sky exposure has doubled. In addition to this increase in exposure, the entire dataset recently underwent a full reprocessing, applying refined instrumental calibrations and event selections. This recalibration, which followed procedures as described in (Atwood et al. 2013), lead to significant improvement in sensitivity and source localization and resulted in the FL8Y source list used here.

We have updated the relation between \gray luminosity and core radio luminosity using our expanded data set as discussed in Section \ref{newsources} in order to derive an new core radio to $\gamma$-ray luminosity relation. Using our new data set we find a best fit to the relation shown in Fig. \ref{gamma-radio} to be of the form of equation (\ref{lum-lum}) with derived values of $A = 0.89$ and $B = 7.15$.
Figure \ref{gamma-radio} shows the luminosity data overlaid with our derived correlation. The previous results of di Mauro et al.(2014) and Hooper
et al. (2016), are also shown for comparison in Figure \ref{gamma-radio}.\footnote{\small{Previous results were based on the smaller (3FGL) sample with about half the number of detections, and both incorporated upper limits in to their analysis. In calculating the parameters of the luminosity correlation, we have not included sources with only \gray flux
upper limits. However, our result agrees well with that of di Mauro (2014).}}

\section{Low Luminosity Core Dominated Radio Galaxies}
\label{lowlum}

In roughly the past decade populations of low luminosity compact core dominated radio sources have been revealed through a combination of wide field-of-view $\sim$ GHz surveys and follow up interferometric mapping. These abundant sources have been designated "miniature RGs" or "Core Galaxies" (CoreG) with typical luminosities in the range $10^{20} - 10^{22}$ W/Hz at 1.4 GHz  and "FR0 galaxies" with luminosities in the range $10^{22} - 10^{24}$ W/Hz at 1.4 GHz (Balmaverde \& Capetti 2006). These highly core-dominated radio sources are found mostly in luminous, red, early type galaxies (Herrera Ruiz et al. 2017; Baldi, Capetti, \& Massaro, 2018). They are classified predominantly as LERGs (low-excitation radio galaxies) (see, e.g. Best \& Heckman (2012a).
FR0 galaxies, although of lower radio luminosity than FRI galaxies, have been found to be five times more numerous than FRI galaxies in the local universe (Baldi et al. 2018). 

Core G and FR0 galaxies are faint radio sources, having luminosities below $10^{25}$ W/Hz. Because of their relatively low luminosity, they are expected to be extremely difficult to detect as individual \gray sources, as follows from relation (\ref{lum-lum})\footnote{\small{There has been one claimed detection of an FR0 designated as Toll 326-379 as a \gray source (Grandi, Capetti \& Baldi 2016). However, with the refined positional determination of this gamma-ray source in the FL8Y source list the association with Toll~1326-379 is now very questionable.}}. In fact, even at radio frequencies these galaxies have not been observed at redshifts above $\sim$ 0.5. However, owing to their abundance, we consider core-dominant Core G and FR0 galaxies to be candidates to contribute significantly to the unresolved extragalactic \gray background.  

We have searched the literature as well as the Sloan Digital Sky Survey (SDSS) 
database for optical spectroscopy of the {\it Fermi}-detected radiogalaxies. The majority
of them have an available optical spectrum which allows us to classify $\sim$93\% of these radiogalaxies as either "HERGs" (high excitation radio galaxies) or "LERGs". Whereas HERGs typically have accretion rates of $\sim$ 1-10 \% Eddington, LERGs have typical accretion rates below $\sim$ 1 \% Eddington (Heckman \& Best 2012). Using their optical spectra to classify the radio galaxies detected by {\it Fermi}, we classify
$\sim$70\% of them as LERGs and $\sim$30\% as HERGs. The later are usually indicated by their broad
permitted lines, which make them either quasars or Seyfert 1 AGN.

Baldi et al. (2018) chose their sample of 108 FR0 galaxies to be limited to redshifts $z \le 0.05$ to optimize spatial resolution, and to fluxes $F_r \ge$ 5 mJy. The Baldi et al. FR0 sample peaks in the range 20 mJy $\ge F_r$ 30 mJy. Given their resolution, they find their observed FR0s to be very strongly core dominated. 

We have examined optical spectra of the FR0 sources which have available
observations from SDSS. Of those, somewhat more than 80\% are likely 
LERGs. Their population is therefore likely to show the relatively weak redshift dependence
typical of LERGs (Best, et al. 2014; Pracy et al. 2016; Yuan et al. 2017), in contrast with the strong redshift-evolution seen for HERGs.
\footnote{Given the LERG predominance among the FR0 sample, we note that the converse is not necessarily true, i.e., one cannot conclude LERGs as a population are not predominantly FR0s.} 

Since FR0s are low redshift sources, with an expected weak redshift dependence, we make the reasonable conservative assumption of a Euclidean source count for these sources. This assumption is supported by the fact that FR0s are predominantly LERGs. It is known that LERGs have little redshift dependence (e.g., Pracy et al. 2016). 

Figure \ref{Euclid} shows the integral source count distribution obtained by Baldi et al. (2018) for the 108 FR0 galaxies in their sample. A Euclidean distribution extrapolation is also shown in the figure. Thus, it would be difficult to obtain the contribution of {\it FR0s alone} to the EGB from this sample. 

We wish to estimate the possible contribution of {\it all} core dominated radio sources to the EGB based on a more complete picture. This would include low luminosity sources such as FR0s, but also higher luminosity core dominated sources. Our method for doing this will be discussed in the next section.

\section{Luminosity Function (LF) of Radio Core Emission}
\label{LF}

Traditionally galaxy radio luminosity functions are derived based on
the total; i.e. spatially integrated or, more commonly unresolved
radio emission. In recent years combined radio and spectroscopic
surveys have led to statistically complete samples comprising $\sim
10^4$ objects (e.g., Best \& Heckman (2012); Pracy et al. 2016) from
which rigorously determined radio luminosity functions have been
derived, but most of these sample objects are unresolved.  

As discussed earlier, the core radio
emission in radio galaxies is what is physically most directly associated with the
central AGN activity. As such a luminosity function based on a large,
flux limited statistically complete sample of resolved radio galaxy
core luminosities would be greater value for many applications
involving AGN and related high-energy properties such as their
\gray emission and contribution to the EGB. In practical terms
however, the task of obtaining VLBI maps for a large sample of say
$10^3$ to $10^4$ galaxies is a daunting one, and thus progress has
remained limited. The previous efforts to estimate the radio-galaxy
contribution to the EGB have invoked scaling relations to estimate the
radio core luminosity function from the total one (e.g., Fig 4 of di Mauro et
al. 2014), however relations have necessarily been based on limited
statistics. Thus, uncertainties are inherent in estimates of the EGB.

Recently, Yuan et al (2018) have made progress in towards improving
estimates of the core radio luminosity function by employing a large
sample of radio galaxies in conjunction with statistical methods. They
find a relatively weak luminosity dependent evolution in comparison to
other derivations of the total luminosity functions. We note that their 
derivation incorporates both FRI
and II subclasses which are delineated in terms of luminosity, but the
total-to-core correction for the combined
luminosity distribution is smoothly interpolated. The sample also extends
to the low-luminosity extreme core dominant FR0 population, which may in fact be the
predominant population locally (Baldi, Capetti, A. \& Massaro,
2018). The resulting radio galaxy core luminosity function has thus been 
obtained by Yuan et al. (2018). Yuan et al. also show that relatively {\it low luminosity, 
low redshift radio sources below $10^{25}$ W/Hz are core dominated.} This luminosity function at low luminosities is consistent with the Euclidean number flux distribution shown in Figure \ref{Euclid}.

We thus chose to compute our estimate of 
the radio galaxy contribution to the EGB based on the Yuan et al (2018) luminosity
function. We use the LFs for sources at higher redshifts as also derived by Yuan et al (2018) in Figure 5 of their Section 5. The higher redshift LFs derived by Yuan et al. show a very weak redshift evolution as typical of LERGS, with the number density peaking at $z \sim 0.8$. This resonates with our results in Section \ref{lowlum} where we found that both FR0s and \gray-loud radio galaxies are primarily LERGs.

\section{Gamma-ray Background from Core Dominated CoreG, FR0 and FRI Galaxies}
\label{EGB}

We proceeded by deriving a \gray luminosity function based on our
$L_\gamma$ to $L_R$ correlation presented in section \ref{rl-gl} and the radio
galaxy core luminosity function of Yuan (2018) as discussed in Section \ref{LF}.
Specifically, we have used the parametric form of that LF and their
fitted parameters for the radio galaxy population. From that we
compute the \gray LF as 
\begin{equation}
\Phi_{\gamma}(L_\gamma, z) = \Phi_R(L_R(L_\gamma),z) d\log L_R d\log L_{\gamma},
\label{GR}  
\end{equation}
with equation (\ref{GR}) being evaluated using the relation (\ref{lum-lum}), extrapolated to lower luminosities than were detected by {\it Fermi}. Here and elsewhere
$L_R$ refers to the core luminosity at a frequency of 5 GHz. We then compute the EGB
contribution at a given energy according equation ({\ref{eq:grb}), except that we
have not included the photon-photon opacity term. Also, we have
included a point source detection efficiency term which effectively
removes the contribution from the brightest sources. Such sources
would presumably be above the point-source threshold such as the 45
detections used in our gamma-to-radio luminosity relation and would be
masked in the empirical EGB measurement process. For this procedure we
follow the methodology of Abdo et al.(2010).

We now calculate the \gray background from core dominated radio galaxies. We assume that these sources have \gray spectra of approximate power-law form $KE^{-\Gamma}$ in the energy range $0.1 < 100$ GeV. We assume a distribution of \gray spectral indexes, $\Gamma$, for the 45 non-blazar radio loud AGN taken from the {\it Fermi} FL8Y
compilation, fit to a Gaussian distribution with $\sigma$ = 0.35 around a mean value of 2.3. This is shown in Figure \ref{SID}. Our Gaussian assumption produce a slightly concave \gray spectrum (e.g., Stecker \& Salamon 1996). We note that, despite the limited statistics, this our mean value matches precisely the measured EGB slope over the nominal 1-100 GeV range (Ackermann et al. 2015).

We use the luminosity function $\Phi_{R}(L,z)$ taken from Yuan et al. (2018)
together with equation (\ref{GR}) 
to obtain the \gray luminosity function for core dominated radio galaxies. The \gray background is then given by 
\begin{equation}
\label{eq:grb}
\begin{array}{c}
I(E_\gamma) = \int_{\Gamma_{min}}^{\Gamma_{max}} \frac{dn(\Gamma)}{d\Gamma} d\Gamma \int_0^{z_{\mathrm{max}}} dz 
\int_{L_{\gamma,
\mathrm{min}}}^{L_{\gamma, \mathrm{max}}} \frac{dL_\gamma}
{\ln(10) L_\gamma} \times \\
\\
\frac{d^2 V}{d\Omega
dz}  \Phi_{\gamma,Core}(L_\gamma, z) \frac{dF_{\gamma}(L_\gamma,
(1+z)  E_\gamma, z)}{dE_\gamma}\ e^{-\tau(E_\gamma, z)}\ 
\end{array}
\end{equation}

In order to calculate the EGB from core dominated radio galaxies,
we carried out the integration of equation (\ref{eq:grb}) over the range
of observed spectral indices and with limits of integration in
redshift and luminosity that covered the pertinent
ranges. The \gray luminosity range used, viz. 0.1 - 300 GeV, corresponds
to the range of radio galaxy core luminosities (in erg/s) $36.0 < \log L_{\gamma} < 45.0$ (see Figure \ref{lum-lum}).  We evaluated the comoving volume
integration from redshift $z = 0$ to $z = 3.0$. 

Aside from our basic assumption regarding the low-luminosity
core-dominant radio galaxy population as $\gamma$-ray emitters, the
largest source of uncertainty in our calculation is the inherent
scatter in the $\gamma$-ray-to-radio luminosity correlation as given by 
equation (\ref{lum-lum}). This scatter is shown in Figure \ref{gamma-radio}. To
quantify this scatter, we considered a parallel band of uncertainty about the red
curve of Fig. \ref{gamma-radio} spanning a factor of 2 (or 0.3 dex). We then computed
the EGB, separately varying the parameters A and B in equation (\ref{lum-lum}), with the
curves constrained to lie within this certainty band. By comparison, the uncertainty bands of the Yuan et al (2018) luminosity functions employed in our calculation have uncertainty bands less than 0.1 dex in vertical extent at most and less than 0.05 dex near the knee where most of the contribution to the total luminosity occurs. Thus, the uncertainty in the luminosity function comprises a much smaller contribution to our uncertainty than the scatter in the gamma-ray to radio correlation.  

Taking these uncertainties into account, we find that our evaluation using equations (\ref{lum-lum}) - (\ref{eq:grb}) gives a range of flux values from $\sim$ $11\% \pm 7\%$ of the observed EGB between 100 MeV and 100 GeV. This is shown by the pink uncertainty band depicted in Figure \ref{DGRB}. Our results thus indicate that the total \gray flux from core dominated radio galaxies can account for a minor, but non-negligible, $11 \pm 7\%$ of the unresolved \gray background observed using the {\it Fermi}-LAT telescope.

\section{Conclusions}
\label{concl}

Using radio maps we determined estimated core radio fluxes for the \gray emitting radio sources given in the expanded {\it Fermi} FL8Y source list. We used those results, to derive a new log-linear relation between \gray luminosity and core radio luminosity for {\it Fermi}-detected radio galaxies. We then considered the importance of the potential contribution of the recently studied classes of low luminosity radio sources to the diffuse GRB; such sources being too weak to be individually detected by {\it Fermi}. Using luminosity functions for core dominated radio galaxies at different redshifts, including the classes of low luminosity sources discussed in Section \ref{lowlum}, we determined the contribution to the EGB from unresolved radio galaxies. 
 
In comparing our results with the unresolved EGB derived by the {\it Fermi} collaboration (referred to by Ackermann et al. 2015 as the IGRB), we note that the spectral index of the observed background is consistent with $E^{-2.3}$ so that the fraction of the background that we may attribute to core dominated radio galaxies is roughly independent of energy. However, the unresolved background derived by Ackermann et al. (2015) has a steep drop off above $\sim$ 350 GeV. This drop off is probably due to a combination of (a) decreased sensitivity at the highest energies, (b) uncertainty in the analysis of the galactic foreground, and (c) some possible absorption from interactions with EBL photons (e.g., Stecker, Scully and Malkan 2016). For this reason, we do not calculate our results for energies above $\sim$ 100 GeV. In this regard, we note that the {\it Fermi} group has determined that $\sim$ 85\% of the unresolved \gray background above 50 GeV is from blazars (Ackermann, M. et al. 2016).

We note that the {\it Fermi} upper limits on non-detected radio galaxies (Di Mauro et al. 2014) are consistent with relation (\ref{lum-lum}) within our error band.
Thus, while it is possible that not all core-dominated radio sources are \gray sources,
our hypothesis and conclusions are reasonable. 

On the other hand, we also note that the LF that we have used, taken from Yuan et al. (2018), may underestimate the number of lower luminosity radio sources, e.g., FR0s, owing to incompleteness. The LF below the knee of the LF may actually be steeper than assumed (see e.g. Falcke, K\"{o}rding \& Nagar 2004). 

Our results, shown in Figure \ref{DGRB}, show that the total \gray flux from core dominated radio galaxies can account for a minor, but non-negligible, $4\% - 18\%$
of the \gray background over the main energy range observed by {\it Fermi}. Together with the contributions from blazars and star forming galaxies (e.g., Stecker \& Venters 2011), the entire observed diffuse \gray background can reasonably be accounted for.

\begin{deluxetable}{cccccc}
    \tablecolumns{7}
\tabletypesize{\scriptsize}
\tablewidth{0pt}
\tablecaption{{\bf Radiogalaxies Detected in 4FGL}\label{table:obs}}
\tablehead{ &  &
  \colhead{$\rm S_{core}$} & 
  \colhead{} & &
  \colhead{} \\ 
  \colhead{Fermi ID} & 
  \colhead{Source ID} & 
  \colhead{{(Jy})} & 
  \colhead{Ref.\tablenotemark{*}} & 
 \colhead{Tel/Array} & 
  \colhead{(HERG/LERG)} 
 \\ }

\startdata
FL8Y J0030.6-0212 &  SDSS J002900.98-01134  & 0.23 &  Lee17  & ATCA 20 GHz   & LERG \\  
FL8Y J0057.7+3022 &  NGC 315  & 1.4 &  Mack97  & Effelsberg  & LERG  \\  
FL8Y J0308.4+0407 & MRC 0305+039 = NGC 1218  & 0.7 &  Unger84  & Merlin & HERG  \\  
FL8Y J0316.9+4120 & IC 310  & 0.17 &  Ahnen17  & VLA  & LERG  \\  
FL8Y J0319.8+4130 & NGC 1275  &  3--28  &  Condon86  & VLA  & HERG  \\  
FL8Y J0322.6-3712e & MRC 0320-373 = NGC 1316 = Fornax A  & 0.12 &  Fomalont89  & VLA & LERG  \\  
FL8Y J0334.3+3920 & 4C 39.12  & 0.43 &  Liuzzo13 & VLA-8GHz & LERG  \\  
FL8Y J0418.5+3810 & 3CR 111  & 1.2 &  Hardcastle88  & VLA  & HERG  \\  
FL8Y J0433.0+0522 & MRC 0430+052 = 3C 120  & 1 &  Fey96   & VLBA \ & HERG  \\  
FL8Y J0519.7-4546 & Pks 0518-45 = Pictor A  & 15.4 &  Tingay03   & ATCA 5 GHz  & HERG  \\  
FL8Y J0521.2+1637 & 3C 138  & 0.88 &  Akujor91   & MERLIN \ &  HERG  \\  
FL8Y J0522.9-3628 & MRC 0521-365  & 1 &  Lee16   & VLA & HERG  \\  
FL8Y J0524.7-2405 &  MRC 0522-239  &  $< $0.5   &  Kapahi98  & VLA  &  ?  \\  
FL8Y J0623.9-5300 &  MRC 0620-526  & 1.2 &  Chherti13 & ATCA 20 GHz & LERG  \\  
FL8Y J0627.1-3529 &  MRC 0625-354  & 0.5 &  Morganti97   & VLA  & LERG  \\  
FL8Y J0708.9-3616 &  MRC 0707-35  & 0.045 &  Saikia99;Yuan+Wang'12  & ATCA 5 GHz &  ?  \\  
FL8Y J0758.7+3746 &  3C 189 = NGC 2484  & 0 &  Condon91    & 5 GHz & LERG  \\  
FL8Y J0840.8+1315 &  3C 207  & 0.35 &  Kharb10   & VLA & HERG  \\  
FL8Y J0858.3+1409 &  3C 208  & 0.51 &  Bridle94  & VLA & HERG  \\  
FL8Y J0934.3+3926 &  3C 221  &  $<$0.4   &  Vigotti99   & 5 GHz  &  ?  \\  
FL8Y J1040.5+0617 &  SDSS J103928.21+05361   & 0.2 & Kozei\l 11 & 1.4 GHz & LERG    \\  
FL8Y J1053.6+4930 &  SDSS J105344.12+49295  & 0.3 &  vanVelzen15   & VLA  & LERG  \\  
FL8Y J1144.9+1937 &  3C 264 = NGC 3862  & 0.4 &  Bridle81   & VLA  & LERG  \\  
FL8Y J1219.5+0548 &  3C 270 = NGC 4261  & 0.3 &  Jones97   & VLA  & LERG  \\  
FL8Y J1230.8+1223 &  3C 274 = M 87  & 3.9 &  Kharb10    & VLA  &  LERG \\  
FL8Y J1244.5+1616 &  3C 275.1  & 0.2 &  Fernini14   & VLA  & HERG  \\  
FL8Y J1312.6+4828 &  SDSS J130248.70+4755105  & 0.033 &  Best12b  & VLA  & LERG  \\  
FL8Y J1306.4+1112 &  SDSS J130619.24+11133  & 0.2 &  Lin10  & VLA  & LERG  \\  
FL8Y J1306.7-2149 &  MRC 1303-215   & 0.1  &  Kapahi98  & VLA &  LERG \\  
FL8Y J1325.5-4301 &  MRC 1322-427 & 7 &  Yuan12   & VLA  & HERG  \\  
FL8Y J1331.0+3031 &  3C 286  & 5.5 &  An04   & VLA & HERG  \\  
FL8Y J1346.3-6027 &  Cen B  &  2--6  &   Chherti13  & ATCA  &  LERG  \\  
FL8Y J1350.8+3033 &  3C 293 = UGC 8782  & 1.55 &  Joshi11   & VLA & LERG  \\  
FL8Y J1428.5+4240 &  SDSS J142832.60+42402  & 0.2 &  Liuzzo13   & VLA & LERG  \\  
FL8Y J1443.0+5201 &  3C 303  & 0.13 &  Hardcastle88    & VLA  & HERG  \\  
FL8Y J1454.2+1622 &  3C 306 = IC 4516  & 0.002 &  Condon91   & VLA & LERG  \\  
FL8Y J1459.0+7139 &  3C 309.1  & 2.4 & Yuan18    & VLA  & HERG  \\  
FL8Y J1518.7+0202 &  SDSS J151640.21+00150   & 0.128 & Lister16    & VLBA  & HERG  \\  
FL8Y J1518.6+0613 &  SDSS J151845.72+06135  & 0.1 &  Owen92   & VLA & LERG  \\  
FL8Y J1521.2+0421 &  Parkes 1518+045 = CGCG 049-138  &  &  Duncan93    & VLA  & LERG  \\  
FL8Y J1631.0+8233 &  NGC 6251  & 0.24 &  Pushkarev12    & VLA  & LERG  \\  
FL8Y J1829.5+4846 &  3C 380  & 4.5 &  Yuan12  & VLA  & HERG  \\  
FL8Y J1843.5-4835 &  MRC 1839-487  & 0.16 &  Yuan12  & VLA  & LERG  \\  
FL8Y J2004.1-3603 &  MRC 1955-357  &  $<$ 0.21  &  Chherti13  & ATCA 20 GHz  &  HERG  \\  
FL8Y J2329.7-2117 & MRC 2327-215  &  $<$ 0.1  &  Kapahi98   & VLA  &   LERG
\enddata

$\tablenotemark{*}$ See reference list for full references.
\end{deluxetable}

\begin{figure}
\includegraphics[width = 7.0 in]{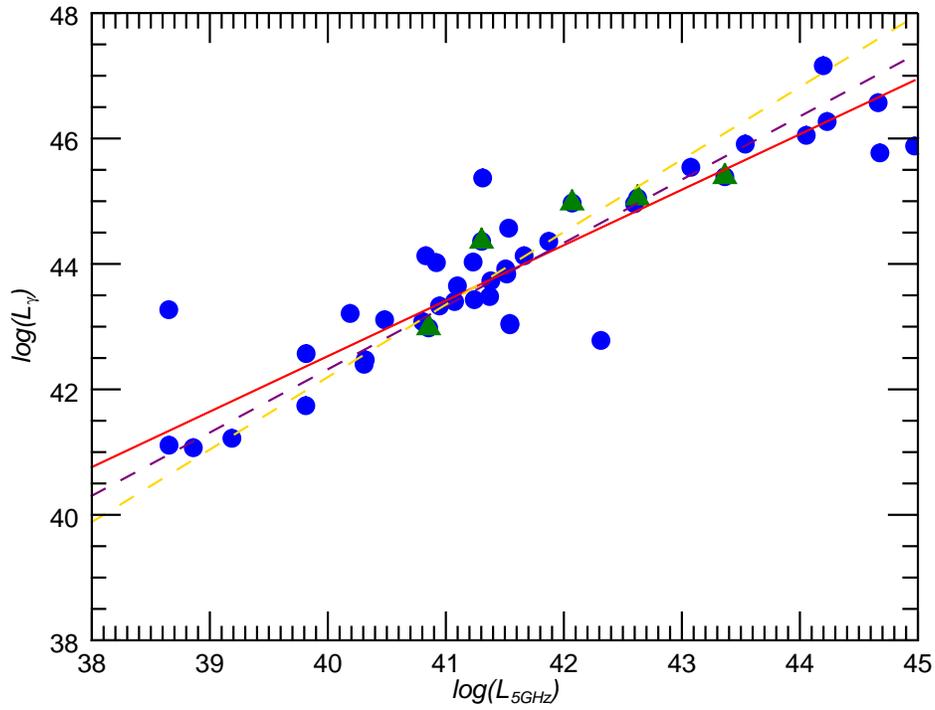}
\vspace{-6.0cm}
\caption{Gamma-ray luminosities of radio galaxies from the {\it Fermi} FL8Y source list vs. estimated core radio luminosities at 5 GHz. All luminosities are given in $erg/s$. Blue circles denote LERGs and unclassified sources; green triangles denote HERGs. The red line is our fit (see eq. (\ref{lum-lum}). Previous results derived using the earlier 3FGL {\it Fermi} catalog are also shown by dashed lines for comparison (Purple:Di Mauro et al. 2014, Yellow:Hooper et al. 2016).}
\label{gamma-radio}
\end{figure}

\begin{figure}
\includegraphics[width = 7.0 in]{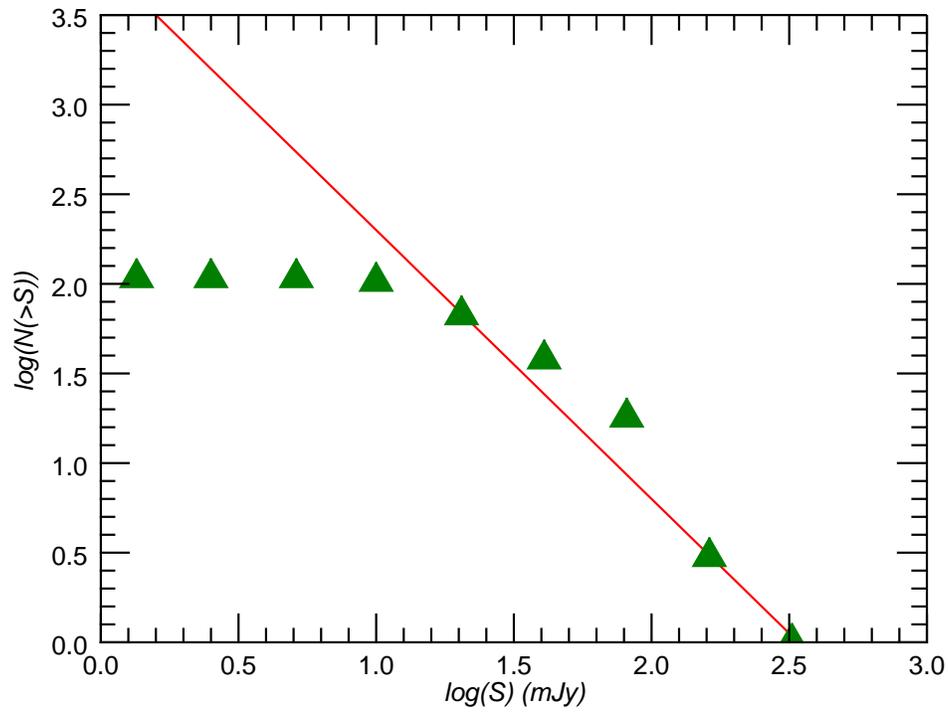}
\vspace{-6.0cm}
\caption{Flux limited FR0 number-flux distribution from Baldi et al.(2018). The plot is for the 108 Baldi et al. FR0 galaxies covering one quarter of the sky. The sample appears to become incomplete below $\sim$ 30 mJy (see text). 
We also show our assumed Euclidean distribution extrapolated to 1 mJy.}
\label{Euclid}
\end{figure}

\begin{figure}
\includegraphics[width = 7.0 in]{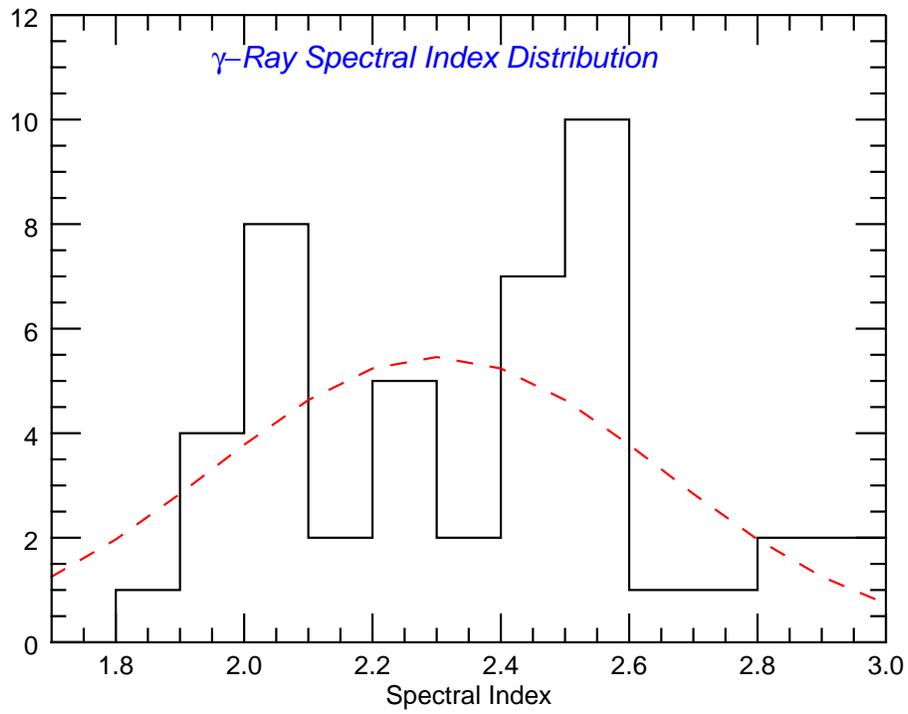} 
\vspace{-6.0cm}
\caption{Distribution of \gray spectral indexes for the 45 non-blazar radio loud AGN taken from the {\it Fermi} FL8Y
compilation. The red dotted curve is the best fit Gaussian distribution with a mean value of 2.3 and a width of 0.35.}
\label{SID}
\end{figure}

\begin{figure}
\includegraphics[width = 7.0 in]{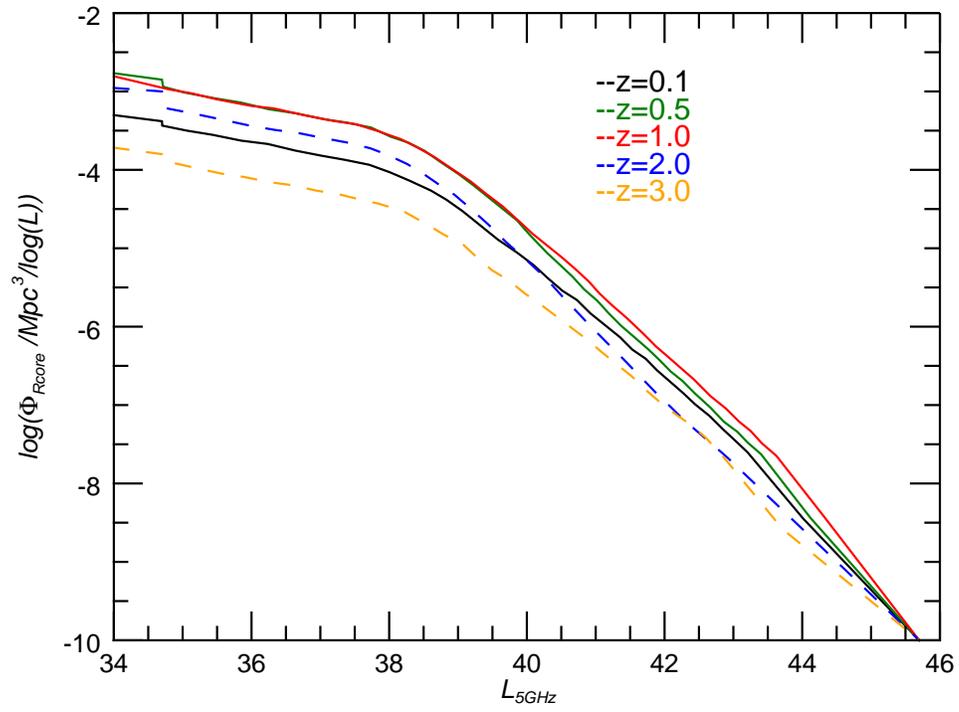}
\vspace{-6.0cm}
\caption{Radio galaxy core luminosity functions at 5 GHz derived from Yuan et. al. (2018) and used in our calculations as described in the text. The horizontal axis units are in $ergs/s$.}
\label{coreLF}
\end{figure}

\begin{figure}
\includegraphics[width = 5.0 in, angle=270]{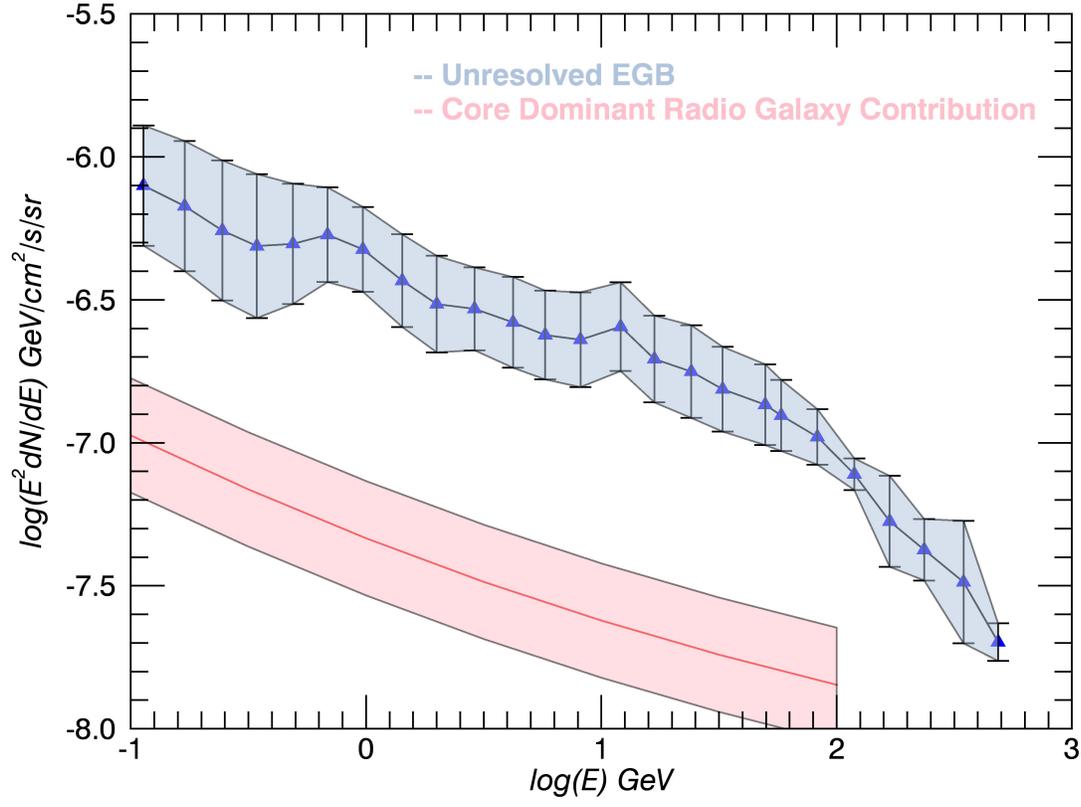} 
\caption{Fermi measurements of the unresolved extragalactic \gray background (Ackermann et al. 2015) are shown in blue.  Results of our calculation are shown by the red line and pink uncertainty band (see text). We find that $11\pm7$\% of the unresolved EGB at $\sim 1$ GeV can be due to unresolved, core dominant radio galaxies. }
\label{DGRB}
\end{figure}

\end{document}